\documentclass[12pt]{article}
\usepackage{geometry}                
\usepackage{graphicx}
\usepackage{amssymb}
\usepackage{epstopdf}
\def\0#1{{\mathrm{#1}}}
\def\1#1{{\mathbb{#1}}}
\def\2#1{{\mathbf{#1}}}  
\def\3#1{{\mathcal{#1}}}
\def\4#1{{\mathsf{#1}}}   
\def\5#1{{{\widetilde{#1}}}}  
\def\6#1{{\overline{#1}}} 
\def\7#1{\breve{#1}}
 \def\8#1{{\widehat{#1}}}  
 \def\9#1{{\widecheck{#1}}}
 
 \def\cir#1{{\stackrel{\circ}{#1}}}

 \def\<{{\left<\right.}}
\def\>{{\left.\right>}}

\def\x{\times}

\def\BAR{\begin{array}}
\def\EAR{\end{array}}
\def\BEQ{\begin{equation}}
\def\EEQ{\end{equation}}
\def\BEN{\begin{enumerate}}
\def\EEN{\end{enumerate}}
\def\BIT{\begin{itemize}}
\def\EIT{\end{itemize}}
\def\BEA{\begin{eqnarray}}
\def\EEA{\end{eqnarray}}
\def\BED{\begin{description}}
\def\END{\end{description}}
\def\BET{\begin{table}}	
\def\ENT{\end{table}}

\def\apo{\mbox{\bf '}}
\def\cliff{\mathop{{{\sqcap}}}\nolimits}
\def\Cliff{\mathop{{\mathsf{Cliff}}}\nolimits}
\def\grade{\mathop{{\mathsf{grade}}}\nolimits} 
 
\def\dim{\mathop{{\mathsf{dim}}}\nolimits} 

\def\dup{\mathop{{\mathsf {dup}}}\nolimits} 

\def\op{{\mathop{\mbox{op\,}}\nolimits}}

\def\oar{\rightharpoonup}	

\def\rao{\leftharpoonup}	

\def\Qi{{\cir{\imath}\,}}
\def\so{\mathop{{\mathsf {so}}}\nolimits}

\def\spin{\mathop{{\mathsf {spin}}}\nolimits}

\def\W{{\bigwedge}}
\def\w{\mathop{\wedge}\nolimits}

\newtheorem{proposition}{Proposition}
\newtheorem{hypothesis}{Hypothesis}

\title{Quantum field theory in quantum set algebra}

\author{David Ritz Finkelstein\\
\small Georgia Institute of Technology, Atlanta, Georgia\footnote{Emeritus.}\\
\small finkelstein@gatech.edu}

 \date{}  

\begin{document}
\maketitle

\begin{abstract}
A modular quantum architecture is given for the space-time,  particles, and fields of the Standard Model and General Relativity.
It assumes a right-handed neutrino, so that  based on their multiplet structure all fundamental fermions have isospin 1/2. 
This opens the possibility that the Higgs field can be identified with the Yang $i$-field of 1947.
The quantum gravitational metric form proposed is a quantification of the Killing form of  the  quantum space-time cell.
There is no trace of the black hole phenomenon at the one-cell quantum level.
\end{abstract}

\section{On quantized space-time\label{S:X}}

This section title pays homage to Snyder and Yang
\cite{SNYDER1947, YANG1947}, 
who turned attention from the symmetry algebra
of space-time to the dynamical algebra of a quantum in space-time,
and regularized it
in an attempt at a regular quantum field theory.
Here the Yang regularization is extended from the abstract algebra to its representation,
within the multiordinal Clifford algebra $\3S$
of \cite{FINKELSTEIN2014},
and a regular quantum field theory
is outlined.

The canonically conjugate orbital variables $x^m, p_m$
are derived here from basic Clifford spin variables $\gamma^n$
much as
Bose statistics is derived from odd statistics.

Yang \cite{YANG1947} reformed the orbital algebra
\BEQ
\4a_{\mbox{\scriptsize orb}}:=\4a(x^{\mu}, p_{\mu}, L_{\mu'\mu},i)
\EEQ
of position, momentum, angular momentum, and $i=\sqrt{-1}$
to make it simple.
Here all these operators have been made anti-Hermitian
with factors of $i$ where necessary.
The reform of $\4a_{\mbox{\scriptsize orb}}$ is
the algebra $\so(5,1)$ of angular momenta $J_{m'm}$ of a point particle
in a space $6\1R$ with dimensionless real coordinates $\eta^m$
with invariant quadratic form
\BEA
d\tau^2&=&g_{m'm}d\eta^{m'}d\eta^m,\cr
g_{00}&=&-1,\cr
g_{11}&=&g_{22}=g_{33}=g_{44}=g_{55}=+1,\cr
J_{m'm}&=&\eta_{m'}\partial_m-\eta_{m}\partial_{m'}.
\EEA

Introduce quantum units of time $\4T$ and energy $\4E$,
and adopt $\hbar$ and $c$ as units of action and speed.
Yang's reformed orbital variables can be rewritten as
\BEA
\cir x{}^{\mu}&=&\4T J^{6\mu},\cr
\cir p_{\mu}&=&\4E J_{5\mu},\cr
\cir L_{\mu'\mu}&=&J_{\mu'\mu},\cr
\Qi &=&\4E\4T J_{65}.
\EEA
Yang imposes a de Sitter subsidiary condition
\BEQ\label{E:DESITTER}
[-x_0{}^2+x_1{}^2+x_2{}^2+x_3{}^2+x_4{}^2]\psi=-\4E^{-2}\psi.
\EEQ
The  Yang $x^1, x^2, x^3$ again have (imaginary) discrete spectra but not $x^0$.

To centralize $\Qi$ requires that the angular momentum $J$ be polarized
so that
$J_{65}$ has a maximum magnitude $j$ in the vacuum:
\BEQ
(J_{65})^2\doteq -j^2, \quad j\gg 1.
\EEQ

The usual condition $i^2=-1$ requires that
\BEQ
j\4E\4T=1\/.
\EEQ

The Yang orbital commutation relations are a special case
of the {\em simple quantization} relations
\BEQ\label{E:C}
[q^{''}, q^{n'}]=C^{n''n'}{}_n q^n
\EEQ
where $C$ is the structure tensor of a simple Lie algebra $\4a$.
In the Yang case $\4a=\so(5,1)\oar \4a_{\mbox{\scriptsize orb}}$
and the $q$'s are 15 orbital variables.
Canonical quantization is a singular limit of simple quantization
as Bose-Einstein quantification is a singular limit of Palev quantification.
Palev quantization is an inverse of 
the composite process consisting of a Palev quantification 
with algebra $\4a$ 
followed by a singular limit $J_{65}{}^2\to \infty$
as the other components of $J$ remain finite.

Thus the proposed simple quantum space-time 
is composed of finitely many  
spin-pairs of rank 5 having Palev statistics.

The general Palev quantization or quantification has 
the commutation relations  (\ref{E:C}) 
of a regular (semisimple, stable)
Lie algebra.  
A Lie algebra that is semisimple but not simple
arises when there are superselection laws between two subspaces
of the port vector space.
Such a limit to superposition 
implies that there are classical variables, so that the quantization process is probably not finished.

The orbital variables of the quantized space-time $\3Y$
constructed next
are again scaled components of an angular momentum tensor $J_{n'n}$ for $\spin(3,3)$
(or $\spin(5,1)$),
but now that of the spinor space $\3S(5)$
resulting from iterated quantification of 
the spin of $\3S(3)$.

While $\3T(2)$ has four Clifford generators $\gamma^a$, 
they have the neutral signature 0, not the Minkowski signature 2.
To reform the orbital algebra $\4a_{\mbox{\scriptsize{orb}}}\oar\spin(3,3)$ we need six Clifford units $\gamma^y$ of neutral signature.
 $\3T(3)$ is the lowest rank Clifford algebra in $\3T$ that will do, with 16 units
$\gamma^n$ of signature 0 and an angular momentum tensor 
$J_{n'n}(3)=\frac 12 \gamma_{n'n}$ representing $\spin(4,4)$.

Yang chose the extra two dimensions to be spacelike for obvious reasons.
Since the two extra dimensions are frozen, however,
they can be timelike without disturbing macroscopic causality.
$J_{n'n}$ represents on the particle rank 5
the angular momentum tensor $\gamma_{n'n}$ of the cell of rank 3.

The quantized space-time $\3Y$ is constructed
from six real $\gamma^y$ 
($y=1,\dots,6$) in 
$\Cliff(3,3)\cong\3T(3)$.
I commonly use a frame in which $g_{n'n}$ is diagonal and
\BEA
g_{11}=g_{22}=g_{33}&=&-g_{44}=-g_{55}=-g_{66}=1,\cr
 \{\gamma^{n'},\gamma^n\}&=&2g^{n'n}.
\EEA
Do not confuse this $\gamma^5$ with the usual $\gamma^5$,
now written $\gamma^{4321}$.

Represent the rank-3 orbital angular momentum tensor $\gamma_{y'y}(3)$
on rank-5 by double quantification:
\BEQ\label{E:J}
J_{y'y}(5)= \sum^{(5)}_{(3)} \gamma_{y'y}(3).
\EEQ
Use $c\hbar\4T$ units, and set $\4E=1/\4N$. 
Then the regular orbital variables in $\3T(5)$ are
\BEQ \label{E:X}
\BAR{lrll}
&\cir p_{m} &:= \frac 1{\4N} J^{m5}&:= \frac 1{\4N} \sum^{(5)}_{(3)} \gamma_{m5},\cr
&\cir x{}^{m}&:= J^{m6}&:= \sum^{(5)}_{(3)} \gamma^{m6},\cr
&\Qi&:= \frac 1{\4N} J^{65} &:= \frac 1{\4N}  \sum^{(5)}_{(3)}\gamma^{65},\cr
\mbox{where}\quad&\cir J_{m'm}& =-\cir J_{mm'}&:= \sum^{(5)}_{(3)} \gamma_{m'm},
\EAR
\EEQ
acting on $\3S{(5)}$.
The relation of $\4T$ to the Planck time is still uncertain.
$\4N$ is the number of spins.

Now all variables that are diagonalizable have discrete bounded spectra.
If a variable is antisymmetric, its square has a discrete bounded spectrum.

(\ref{E:X})  stabilizes the Heisenberg Lie algebra
and the Hilbert space:
\begin{eqnarray}
{\sf h} (x^m, p_m, J_{m'm}, i)
&\rao&
\spin(3,3)\cr
&\cong&{\sf a}(\underbrace{\gamma^{21},\dots,\gamma^{65}}_{15})\cr
\3H &\rao& \3S(5).
\end{eqnarray}
The canonical commutation relations between $x$ and $p$
emerge as the singular centralization $\4N\to infty$, $\Qi\oar i$,
a symmetry-reduction of the Clifford algebra $\op \3S(5)$.
The complex Hilbert space of the canonical theory emerges from the real
Grassmann algebra of the simplicial theory.

\section{Green's functions}

Scattering data of standard quantum field theory are found from 
Green's functions given by poorly defined integrals over field-histories,
\BEA\label{E:GREEN}
G(x_1,\dots, x_n)&:=&\frac 1{\3N} \int [d\psi] \exp iS \;\psi(x_1)\dots
\psi(x_n),\cr
\3N&:=& \int [d\psi] \exp iS 
\EEA
where $\psi=(\psi(x))$ is a general field-history, 
 $S=S[\psi]$ is the action functional operator of the field theory,
 and $\3N$ is a normalizing factor.
 Call the skew-symmetric $iS$ the {\em skew action}.
Reforming $G$ is part of reforming the usual quantum field theory.

 The singular integral over $\psi$ was originally regarded as an integral over classical paths.
There are no classical paths in Nature.
Therefore the path integral is not an operational description 
but a relict of the coarser and more singular classical theory. 
Quantum theory does not assign probability amplitudes to classical possibilities but determines new quantum possibilities too.
It is easier to overlook this when the spectra of both the classical and quantum
variables are continuous and the same.

The path integral is sometimes supposed to be a singular limit of a matrix product of many factors close to unity.
But the actual factors in Nature seem to be portations
of elementary quanta. 
These are very far from unity; their trace is 0 instead of large.
One way to integrate over quantum histories
composed of portations of odd quanta
is to replace the formal integral $\int$
in (\ref{E:GREEN}) by a Berezin integral $\int_{\0B}$.

Operationally put, the usual Green's function (\ref{E:GREEN})
is the evaluation 
\BEQ\label{E:GX}
G(x)=D\circ E(x)
\EEQ
of a
history dynamics dual vector 
\BEQ
D:=\exp iS
\EEQ 
 on a history port vector $E=\Psi_n\dots \Psi_1$.
 $D$ represents the system dynamics
 and $E$ represents the quantum experiment.
  In $\3S$ models both are finite-dimensional
 and the contraction is regular.
 In standard gauge theories, non-compact gauge groups introduce new infinities.
One point of this section is that in $\3S$ theories, 
such non-compactness is no problem.

 It seems plausible that the Berezin integral in (\ref{E:GREEN})
 becomes a finite trace in
 the operational formulation (\ref{E:GX})
 when $D$ and $E$ belong to $\3S(5)$.
 I accept this for now
 and proceed to reform the vectors $D$ and $E(x)$ of the standard theory.

\section{Dynamics\label{S:DYNAMICS}}

Consider only spinor fields $\psi(x)$.
Then the Dirac skew action $iS[\psi]$ is an element 
of an infinite-dimensional Clifford algebra $\3C_{\0D}$.
The linear operators on $\3S(5)$ 
form a regular Clifford algebra 
$\cir{\3C}(5) =\3T(5) =\op \3S(5)$.
We may assume that the reform of the skew action functional $iS$ 
is a polyadic $\cir S\in\3T(5)$. 

The factor $i$ in the exponent of $\exp iS$ 
requires special treatment.
It is a singular limit of a quantification
of a angular momentum $\gamma_{65}$,
a component of the angular momentum tensor $\gamma_{y'y}$
in a regular Lie algebra $\cir{\4a}$ that includes both the Yang orbital Lie
algebra
$\spin(3,3)$ and the regularization of the  Standard Model internal
Lie algebra $\4a_{\0{SM}}$ of (\ref{E:ASM}).

{\hypothesis[Regular dynamics] 
There is a regular dynamics $\cir D\oar D$ before the singular limit $\Qi\oar i$, 
and it is invariant under $\cir{\4a}$.}

\vskip8pt
The singular limit $\4a$ includes $x^{\mu}$ among its generators.
Therefore the Standard Model dynamics $D$ is not invariant under $\4a$.
The singular limit must break $\cir{\4a}$ symmetry and leave at least the 
internal and orbital symmetries
$\4a_{\0{SM}}+\4a_{\mbox{\scriptsize P}}$ of the Standard Model and the Poincar\'e group.

Then one immediate reformation replaces 
\BEA
i&\rao& \4N^{-1}J_{m'm},\cr 
S&\rao& {\cir S}{}^{m'm},\cr
iS&\rao& \frac 1{\4N}\0{tr}\, J_{m'm} S^{m'm}
\EEA
where $\cir S{}{m'm}$ is a tensor dual to $J_{m'm}$ in its transformation. 
The polarization in this singular limit makes $i$ dwarf the other components of $\4N^{-1}J_{m'm}$
and reduces the trace  to one term, the usual product.

The core of the usual dynamics is the Dirac skew action
\BEQ
A_{\0D}:=iS_{\0D}:= i\int[dx] \;\6{\psi}(x)\gamma^{\mu} p_{\mu} \psi(x)\/.
\EEQ

The Yang simplification makes the Poincar\'e-invariant Dirac operator $\gamma^{\mu}\partial_{\mu}$ a term in the Yang-invariant
operator $\gamma^{y'y}J_{y'y}$.
$J$ has already been defined by quantification.
It remains to define how the single-cell operator $\gamma^y$ 
acts on 
the general multicell monadic.
Every basic monadic is the association of a
basic polyadic like those of Table I in \cite{FINKELSTEIN2014}.
A glance shows that every basic element 
has a tail of rank 3 on its 
right-hand side, 
representing a spinor of the Yang group.
To act with any $\gamma\in \3T(3)$ on a monadic $e$,
remove the top bar of $e$, act with $\gamma$ on its 
rank-3 spinor tail, and replace the bar.
This operation can be written as 
\BEQ
\gamma=\iota \gamma \6{\iota}.
\EEQ
Here $\6{\iota}$ is a left-inverse of $\iota$
and annihilates basic $g$-adics of grade $g\ne 1$:
\BEQ
\6{\iota}\iota = 1, \quad \6{\iota}\;{\grade}=\6{\iota}.
\EEQ
And $\gamma$ is extended from $\3S(3)$
to $\3S$ by acting only on the rank-3 factor of its argument.

Following Yang, replace the vector factors and $i$ in the 
Dirac action by angular momenta with two extra dimensions.
The result is, up to normalization,
\BEQ
A\rao \cir A=J^{m'' m'}\gamma_{m'}{}^{m} J_{mm''},
\EEQ
an operator on $\3S(5)$ of Clifford grade 6.
Spinor indices of rank 3 on $\gamma$,
of ranks 3 and 5 on $J$, and of rank 5 on $A$ are unwritten.
The Green's function depends on the signature chosen for the metric
$g$ but is not singular in any case.

Scattering port vectors must be defined before scattering amplitudes
can be computed.
The usual scattering ports determine four components of energy-momentum
that do not commute in the $\3S$ theory.
Coherent spin-states can be used instead \cite{LIEB1973}.

\section{Beneath dynamics}

The question of the physical origin of dynamical law
is often raised \cite{WHEELER1983}. 
The dynamical law of the quantum system under study is violated 
at the beginning and end of every experiment,
just as  every symmetry is broken by the experimenter in order to be observed \cite{WIGNER1967}.
Here, as in the Dirac-Feynman summation over quantum histories, the dynamical law is summed up in a high-grade port vector $D$.

Every other port vector in physics represents portation to or from
a statistical population in a reservoir outside the system.
I must suppose that this holds for $D$ too.

The dynamical ``law" is then a statistical correlation
between the system under study and its external source and sink.
Since the ports are coherent (described by vectors rather than statistical operators)
this correlation is an entanglement.
Dynamics of a system is an entanglement with the rest of Nature.

\section{Isospin and color\label{S:ISOSPIN}}

There are 16 kinds
of first-generation fundamental fermion (briefly {\em ffermion},
pronouncable in Welsh) 
in the Standard Model, 4 leptons and 12 quarks;
call them flavors.

Assume that a quark is a lepton with a color;
that is, that there is a module $F=L+Q$ encoding the distinction
among the ffermions,
where the lepton tag L that makes a ffermion a lepton has grade 1, rank 3, multiplicity 4; the quark tag $Q$ has grade $>1$, rank 3, multiplicity 12, and 
$Q=CL$, where $C$ is color of multiplicity 3. 
It seems possible that the strong interactions exchange $C$'s, which leptons do not have.
Leptons do have isospin in the form of so(4R),
acting on the $L$ in the lepton and the $L$ in the quark.

Entire leptons suffer several independent dichotomies:
$\tau_1$ dividing left-handed electrons from left-handed neutrinos, chirality $i\gamma^{4321}$ dividing the left-handed leptons from the right-handed,
polarity $\Pi$ dividing
the lepton imports from the lepton exports,
and particle number  $N\doteq\pm 1$ dividing ffermions proper from anti-ffermions.
These dichotomies cannot act independently on the  lepton tag $L$ alone.
Chirality can act on the spin module of the ffermion;
polarity can act on the entire ffermion. 
Then $\tau_1$ and $N$ act on the four leptons
and also on the 12 quarks.

This classification fits nicely into $\3S(3)$ 
if
the ffermion {\em flavor module} 
has import space $\3F= \3S(3)\cong 16\1R$.
$\3S(3)$ is spanned by the quantum sets with the four-place hyperbinary symbols 
for 0, 1, 2, $\dots$, 15.

Its Clifford algebra has eight flavor generators $\gamma^f\in\3T_1(3)$
($f=1,\dots, 8$).
Take $\gamma^1, \dots, \gamma^4\in \3T_1(3)$ to be left-multiplications by $e^1, \dots, e^4\in
\3S$ (see Table I in \cite{FINKELSTEIN2014}), and 
$\gamma^5, \dots, \gamma^8$ to be the
left partial differentiators with respect to the same four elements.
%
%

Set aside four generators $\gamma^{\mu}$ of the 16 in $\3T(2)$ 
($\mu=1,2,3,4$)
to generate
the Lorentz group; the four in $\3T(1)$ having the wrong signature.
Then extend them with  $\gamma^5,\gamma^6\in \3T(2)$, both
of the same signature,  to six $\gamma^y\in\3T(3)$ ($y=1,\dots,6)$,
to generate $\cir p_{\mu}$, $\cir x^{\mu}$, and $\gamma^{65}\sim \Qi\oar i$ as well.

Since left-handed and right-handed fermions have the same 
multiplicities, assume that both carry isospin 1/2
and that isospin commutes with parity $P$ and chirality,
unlike the Standard Model isospin.

Parity must still be violated.
The isospin of the left-handed pro-particles and their anti-particles
couple to the $W$ meson; that of their mirror images does not.
In $\3S$ models, this means that the $W$ is a pair (or a number of pairs)
of a left-handed odd  pro-particle and its right-handed anti-particle.  
These pre-$W$'s are not necessarily able to exist as free particles,
but since the $W$ couples to both leptons and quarks
the pre-$W$'s should be elements of both.
This chiral isospin coupling correlation must appear in the dynamics vector responsible for the 
binding of fermions into a $W$,
and ultimately in an alignment of the exosystem (the universe outside the system).


The Yang orbital Lie algebra
$\spin(3,3)$
and an internal Lie algebra $\spin(5,5)$
are subalgebras of $\spin(5,5)$.
This differs from GUT $\spin(10)$ in signature.

The two values of isospin, Up and Down, may be attached to the
first two binary places
\BEQ
|U\>= s_1=\;\61 , \quad |D\>=s_2= \;\6{\6{\rule{4pt}{0pt}}},
\EEQ
which are indexed by their serial numbers in Table I. 
Similarly the three values of color may be identified with the three basis elements
\BEQ 
 |R\>=s_4=\;\6{\6{\6{\rule{4pt}{0pt}}}},\quad
 |G\>=s_8=\;\6{\6{\6{\rule{4pt}{0pt}}}\;\6{\rule{4pt}{0pt}}},\quad
  |B\>=s_{12}=\; \6{\6{\6{\rule{4pt}{0pt}}}\;\6{\rule{4pt}{0pt}}}\;\6{\6{\6{\rule{4pt}{0pt}}}},
 \EEQ
 which span a color space $\3C\subset\3S$ such that
 $\3Q=\3L\w\3C$ (quark = lepton $\w$ color).

These five places support the Lie algebra $\so(5)$.
This does not yet contain the GUT $\so(10)$;
it must be doubled.

The Standard Model Lie algebra is
\BEQ\label{E:ASM}
\4a_{\0{SM}}:=\0s[\0u(2)\oplus\0u(3)]
\EEQ 
composed of $5\x 5$ matrices composed of diagonal blocks of $2\x 2$  $\0u(2)$
matrices and $3\x 3$ $\0u(3)$ matrices.
While GUT theories simplify $\4a_{\0{SM}}$ for the sake of unity,
they also regularize it, eliminating hypercharge centrality.
Neglecting signatures for the moment, 
the $\so(10\1R)$ GUT regularization of the Standard Model
and the Yang $\so(6\1R)$ regularization
of the Heisenberg-Poincar\'e group
fit into one $\so(16\1R)$ algebra acting on $\3S(3)$.

\section{Gauge hierarchy \label{S:GAUGE}}

Gauging a fiber Lie algebra $\4g$
creates a bundle Lie algebra $\4G=\4g^{\3M}$ with a local  isomorph of $\4g$
at each point of Minkowski space-time $\3M$,
as in the Standard Model.
The bundle algebra 
is not semisimple, being
 infinite-dimensional.
Therefore it is reformed here. 

Because gaugeons have even statistics, in an $S$ theory they must be an even number
of odd elements \cite{SALLER1974}.                                                                                                                                                                                                                                                                                                                                                                                                                                                                                                                                                                                                                                                         
 

Quantification $\W$ creates copies of the cells, each with a copy  of the cell group.
This suggests the

{\hypothesis [Gauge] 
Gauging is  quantifying a self-dual cell pair and taking a singular limit.
The reform of the fiber gauge algebra is the cell Lie algebra.
The reform of the bundle gauge algebra is the Lie algebra  of the corresponding
subspace of $\3S$.}
\vskip8pt
The reform of the metric and the connection are taken up in Section \ref{S:QG}.

This requires us to represent all the fiber groups of the Standard Model and gravity within
the group of a cell pair.
The least candidate for the port vector space of a cell is 
$\3S(3)$, 
with 16 dimensions.

Roughly speaking,
6 dimensions of the 16 support the Yang orbital algebra $\spin(6)$
and 10 suppport GUT $\spin(10)$.
More precisely, the signatures must fit within the neutral signature of $\3S(3)$; and the two extra Yang dimensions of the 6
may participate in isospin.

The gauge (covariant) differentiator $D$ is, up to a conventional constant $-i\hbar$, the total one-quantum momentum quantified, which in turn is a sector of the quantified angular momentum
\BEQ
J(5)=\sum_{(3)}^{(5)}J(3).
\EEQ

Then the Lie algebra of the cell is
\BEA
\cir{\4G}&=&\grade_2\3T(4)=\spin(\dup \3S(3)),\cr
\3T(4)&=&\op \3S(4)=\cliff \dup \iota\apo\3S(3).
\EEA
$\3T_2(4)$ has a Killing form $\4K(4)$,
a neutral quadratic form on $\3T_2(4)$,
invariant under the gauge group
and unique in this regard up to a numerical factor.

Therefore I propose that this Killing form is the seed for the Einstein form
of quantum gravity.

At the same time, this accounts for the hierarchy of the 
coupling constants.
The Clifford units enter into the orbital momentum (\ref{E:X})
with factors $1/\4N$
and enter into the isospin and color with factors 1.
The rato of the coupling constants is then of order $\4N$.
For two electrons at atomic separations, the ratio of electric to gravitational interaction energy is $\sim 10^{42}$.
This must be about the number of spins in an electron.

\section{Higgs $\phi$, Yang $i$\label{S:HIGGS}}

In the present script the deus ex machina of spontaneous organization is lowered twice, 
first with
the Yang $i$ and then with the Higgs $\phi_{\0H}$.
Perhaps cosmological inflation is also a spontaneous organization,
resulting in another vacuum field, 
a Guth field $\phi_{\0G}$.
Can some of these three organizations be one?

$i$ must have isospin 0 for the dynamics vector
$D=\exp (iS)$
to be
invariant under isospin $\spin(3)$,
and $\phi_{\0H}$ has isospin 1/2 in the Standard Model.
It seems that they cannot be the same.

But it may soon be necessary to update the Standard Model. 
Isospins are historically assigned on the basis of multiplet structure, not interactions.
If the right-handed neutrino exists then the left-handed and right-handed ffermions 
all belong to doublets and must all be assigned isospin 1/2.
I explore this possibility here.

If all the ffermions have isospin 1/2, 
and the Higgs $\phi$ transforms as a ffermion pair, 
it must now have isospin 0 or 1 
instead of the previous value of 1/2.
The term with isospin 1 is indispensable for distinguishing the electric from the weak interactions.
If the pair is symmetric with respect to exchange
of their isospin modules, the isospin 0 part vanishes.
Then the Higgs $\phi$ has three real components, 
as in quaternion quantum mechanics \cite{TAVEL1965},
not two complex ones. 

Parity violation is then concentrated
 in the interaction between the ffermion and the $W$.
In one parsimonious model $W$ transforms as a self-dual pair of only a left-handed ffermion module proper
and its right-handed anti-ffermion, 
without the opposite chiralities.
It need not contain color modules.

Now an invariant dynamical vector $D$ can be formed 
with an isovector $\Qi$
in the form 
\BEQ
D=\exp{\Qi\cdot S},
\EEQ
where both $\Qi$ and $S$ are isovectors, and the exponent is their scalar product.

 To generate isospin $\spin(3\1R)$ within $\3T$
  requires three Clifford monadics $\gamma^{i}$.
  To generate Yang $\spin(3,3)$ requires six $\gamma^y$.
   $i=\phi_{\0H}$ requires two of the $\gamma^i$ to coincide
   with two of the $\gamma^y$.
   Take $y=1,2,3,4,5, 6$ and $i=5, 6, 7, 8$.
   This uses half of $\3T(3)$

 To represent $\spin(3\1R)$ in $\3S$ requires
 four Grassmann monadics $e^5, e^6, e^7, e^8$,
 reserving $e^1, e^2, e^3, e^4$ for Lorentz $\spin(3,1)$.
The isospin 3-vector is then
\BEQ
I_{k}=e^{k'} \tau_{k} \partial_{k'}, \quad \partial_k:=\partial/\partial e^k\/.
\EEQ

As is well known, $\Cliff(4,4)\cong \3T(3)$ is  a  Botts period 
of the real neutral Clifford algebras,
making it a mathematically natural module.
It has a remarkable
triality form
\BEQ
T=\6{\psi}\gamma^{n}p_{n}\psi,\quad \psi\in \3S_-(3),\quad 
p\in \3T(2), \quad \6{\psi}\in \3S_+,
\EEQ
on the three isomorphic
8-dimensional vector spaces 
\BEQ
\3S_+(3) \cong \3T(2) \cong \3S_-(3);
\EEQ
meaning that the value of $T$ on any non-zero vector 
in one of the three spaces is a duality between the remaining two spaces.
Do these beautiful mathematical 
facts have physical meaning?

\section{The 16 fermions\label{S:16FERMIONS}}

The dimension of $\3S(r)$
 grows hyperexponentially
with rank $r$:
\BEQ\label{E:DIM}
\BAR{rcrrrrrrr}
\mbox{for }r&=&0,\; &1, \;&2,\;&3,\;&4,\;&5\cr
\dim \3S(r) &=&1, \; &2,\; &4, \; &16, \;& 64\0K,\;& 2^{64\0K}\cr
\dim \Delta\3S(r) &=&1,\;&1,\;&2,\;&12, \;&65520,\;
&2^{64\0K}-64\0K\;&
\EAR
\EEQ
Rank 5 is the first rank big
enough
for the orbital degrees of freedom of a quantum particle in a quasi-continuous space-time
with chrone comparable to the Planck time.
It is far oversize for the visible universe, but no smaller rank suffices.

If there is a dextral neutrino, 
there is a natural partition of the 
16 kinds of Standard Model first-generation fundamental fermions 
that matches the partition of $\3S(3)$
into four tiers $\Delta \3S(r)$, $r=0,1,2,3$:
\BEQ\label{E:FFERMION}
\BAR{r|rrrrrr}
r & 0& 1 & 2 & 3&  \dots\cr
\hline
\dim \Delta \3S(r)& 1 & 1 & 2 & 12 & \dots\cr
\0{SM\;fermions}&1 & 1 & 2 & 12& 
\EAR
\EEQ
This associates  four leptons with tiers 0, 1, and 2,
and 12 quarks with tier 3.
These tiers do not contain the entire particle,
which ranges up to tier 5 at least, 
but a   module in the particle
that
determines its genus.

Since ranks 0,1, 2, 3, 5, and 6 have plausible interpretations,
probably rank 4 should have one too.
Since it intervenes between quarks and free particles,
one possibility to be considered is that
rank 4 includes the confined assemblies like
the nucleons.
This hypothesis---if it lasts long enough to be dignified by that name---is vulnerable,
since it must agree with the chromodynamic theory of confinement
as far as that agrees with experiment,
but I cannot think of another possible meaning
in current experience for rank 4;
perhaps it has none.

\section{Generation}

Generation-change preserves all the Standard Model symmetries but not the gravitational coupling through mass.
Quantification preserves all the cell symmetries.
This suggests the

{\hypothesis [Generation] 
Generation is a  rank
to which gravitation
couples.}

If all the fundamental quanta indeed have the same rank,
this rank is not the rank of the particle itself
but is presumably the rank of one of its lower-rank modules. 

\section{Quantum gravity\label{S:QG}}

In the gauge theory of gravity, there is a covariant gauge differentiator $D$ that serves as a potential for the gravitational gauge field
$R:=[D,D]$, the curvature.
The gravitational potential $D$ annuls the metric tensor $g$, 
$[D,g]=0$, not by definition but as a constraint in the second-order theory of gravity, or as a dynamical equation in the first-order theory. 
Christoffel solved this for $D$ in terms of $g$ and its Lie derivatives $\partial g$, making $g$ a potential for the potential.
Call any dynamical field whose gauge derivative vanishes, a {\em metroid}, in analogy to the metric.

The Standard Model has a similar pattern of fields and potentials for the non-gravitational gauge fields. Let $D$ include all the gauge potentials, from gravitational to 
chromodynamical, so that the gauge curvature $R$ includes all the gauge fields. The metroids are the metric for gravity and the Higgs field for electricity.

In Dirac spin theory
the local space-time metric is identified with the Dirac anticommutator
\BEQ\label{E:DIRAC}
\{\gamma^{\mu'},\gamma^{\mu}\}=2g^{\mu'mu}\/.
\EEQ
This is a problem for quantum gravity:  $g^{\mu'mu}$  is a classical quantity and
cannot port gravitons. 
When the $\gamma^{\mu}$ belong to one cell,
this $g$ is an unlikely candidate for the seed of the gravitational

After Yang, 
infinitesimal space-time translations are associated with 
infinitesimal rotations $J_{y'y}$ in planes orthogonal to Minkowski space-time.  
The second-grade Clifford elements 
$\gamma_{m 6}, \gamma_{m' 6}$ in
$\3T(4)\cong \Cliff(3,3)$ represent quantum elements of momentum
in the $m$ and $m'$ directions of the cellular quantum space-time
$\3Y$ of Section \ref{S:X}.
Their commutator is
\BEQ\label{E:GGG}
[\gamma_{m5}, \gamma_{m' 5}]= 2\gamma_{mm'},
\EEQ
since $(\gamma^5)^2=-1$.
Since momentum generates translation,
this commutator is an element of curvature in quantum units.

In classical differential geometry the commutator of two such small covariant momenta $P_{m'}, P_m$
would be the angular momentum $J_{m''' m''} R^{m'''m''}{}_{m'}{}^m$, 
where $R$ is the curvature tensor and $J$ is a differential operator
representing an infinitesimal orthogonal transformation.
Here (\ref{E:GGG}) becomes $\gamma_{m''' m''} R^{m'''m''}{}_{m'}{}^m$.

According to (\ref{E:GGG}),
the default scale of magnitude of the effective $R$ is two quantum units.
If quantum units are comparable to Planck units, the quantum unit 
(qu) of curvature is enormous:
\BEA
1 \mbox{ qu(curvature)} &=& 1 \mbox{ qu(length$^{-2}$)} \cr
&\sim& 1/(c\4T)^2\cr
&\sim& 10^{70}\;\0m^{-2}.
\EEA
Yet experimentally the local curvature is $\ll 1\;\0m^{-2}$. 
This means that at the quantum level of resolution $R$ is
a sum of many large terms of both signs that nearly average to 0, like the electric field in a conductor mapped at the resolution 
of the Bohr radius.

According to Riemann's inauguration lecture, 
continuous manifolds have no intrinsic metrical structure while discrete manifolds have an intrinsic metrical structure based on counting steps.
Since quantum manifolds are neither quite continuous nor quite discrete, 
Riemann leaves us in a dilemma.

In fact a
continuous group is a manifold but has intrinsic metrical structure, its Killing form.
Since the orbital variables of a quantum theory generate 
a Lie group, they have its Killing form for a metric.
This is therefore a candidate for the seed 
of the gravitational metric.

In theories of de Sitter and Yang, 
translations are approximations to
rotations about a remote axis.
Therefore they have two vector indices, not one.
For the single cell with port algebra $\3S(3)$,
the orbital elements form the 
angular momentum tensor $\gamma_{m'm}$.

The Killing form is
\BEQ\label{E:K}
K_{\{m'''m''\}\{m'm\}}:= 
\0{tr}\; \Delta \gamma_{m'''m''} \Delta \gamma_{m'm},
\EEQ
where $\Delta$ indicates the adjoint representation,
the difference between left and right multiplication.
This  metric is related to curvature from birth.
For an isolated small cell the center of rotation must be in the cell
and the curvature of the Killing form is large. 

$K$ associates a number $K_{\{m'm\}\{m'm\}}$ 
with the orbital element $\gamma_{m'm}$ as
classical metrics associate numbers with differentials of position.
The quantum metric, however, should associate a graviton port,  
an operator 
in the Clifford algebra
$\3T$, with an orbital element.
The number (\ref{E:K}) could be its expectation value.

Formally, the trace in (\ref{E:K}) is indeed an expectation value, 
namely that of the operator-on-operators
\BEQ\label{E:GDELTA}
g^{\{m'''m''\}\{m'm\}}:=\Delta \gamma^{m'''m''} \Delta \gamma^{m'm}.
\EEQ
This is therefore a better candidate for the seed of the quantum gravitational metric $\cir g$ than $K$.
Call this the Killing operator, 
since its trace is the Killing form.
The quantized gravitational metric is then the quantification
\BEQ\label{E:G}
\cir g^{\{m'''m''\}\{m'm\}}=\2e \;\Delta \gamma^{m'''m''} \Delta \gamma^{m'm}\;\6{\2e}
\EEQ
Like the $g^{\mu'\mu}$ of Special Relativity,
this tensor is Lorentz invariant, in the same sense that
the Dirac vector $\gamma^{m}$ is when the written vector 
and unwritten spinor indices
are consistently transformed.
In the canonical theory  
$g_{\mu'\mu}$, like the tensor $p_{\mu'}p^{\mu}$,
 is symmetric and its components commute,
but not so (\ref{E:G}).

Regular gravitons cannot be exact bosons, 
since the Bose-Einstein commutation relations 
are singular, but
may be palevons that usually pass for bosons.
Then they should obey orthogonal-Lie-algebra 
commutation relations that
have canonical ones as singular limit.
In fact adjoint-formation $\Delta$ is a Lie-algebra isomorphism.
The $ \Delta \gamma_{m'm}$ obey the same commutation relations
as the $\gamma_{m'm}$.
They therefore define the Palev statistics 
of an orthogonal Lie algebra.

At first glance the $g$ of (\ref{E:GDELTA}) does not look like an element of the Clifford algebra
 $\3T$, as $\3S$ observables must be;
but actually it is.
Clifford algebras are closed under $\Delta$:
The left multiplications in $g$ are in $\3T$ by definition, and
the right multiplications can be expressed in terms of
left multiplications and left differentiations, also 
in $\3T$.  
This operator-valued form is then extended from a single cell to macroscopic orbital elements
by multiple quantification.

{\proposition When $\spin(3,3)$ is reduced to the blocks of
space-time $x^{\mu}$, energy-momentum $p_{\mu}$, Lorentz angular momentum $J_{\mu'\mu}$, and
$\1C$,
the Killing form $K$ on $\spin(3,3)$ reduces to the direct sum
of the usual classical physical metrics on these spaces.
} 

These are the Minkowski metrics on space-time and  energy-momentum,
the Lorentz Killing form $J^{\mu'\mu}J_{\mu\mu'}$ on the Lorentz angular momentum,
and the usual $|z|^2$ norm on $\1C$.

\noindent{\bf Proof}. \rule{0pt}{20pt}Direct calculation. 
The Killing form is diagonal in  the basis
$\gamma_{m'm}$, $m'>m$.\kern 5pt$\Box$

As a quick but rough check of this result, 
note that the signatures add up correctly:
\BEQ
2+2+0-1=3=9-6
\EEQ
where the first 2 is the signature of Minkowski space-time,
the second 2 is that of energy-momentum space, 0
is the signature of the Lorentz angular momentum $\|J\|$, 
$-1$ is the signature of $\1C$,
and $3=9-6$ is the signature of the Killing form of 
the 15-dimensional $\spin(3,3)$. 

If all its indices are different, 
$g_{\{m'''m''\}\{m'm\}}$ is symmetric under
\BEQ\label{E:MMMM}
\{m'''m''\}\leftrightarrow \{m'm\}.
\EEQ
This is not the case, however, for the momentum components of a single cell, 
where the fifth Yang dimension occurs twice among the four indices.
Since the momentum $p^{\mu}$ is a quantification of $\gamma^{\mu 5}$
over $\4N$ cells, however,
the pairs exchanged in (\ref{E:MMMM}) 
belong to different cells $\4N-1$ times 
more often than they belong to the same cell,
and symmetry holds when $\4N\to \infty$ even if some indices $m$ repeat.

On the other hand, for finite large $N$, the physical case,
there is a skewsymmetric part to $g$ that is much smaller than the symmetric part, by a factor $\sim 1/\4N$.

For decades Einstein studied the possibility that the electromagnetic field $F$ is the skew-symmetric part $\check g$ of the metric tensor.
Here a skew-symmetric part $\check g$ arises naturally in the quantum theory, which Einstein refused to consider, but 
disappears in the classical limit.
It would violate the Standard Model, in which
$F$ is part of the curvature $[D,D]$, not part of its second potential,
  to identify $\check g$ with $F$. 
Furthermore the  $\check g$ field is as much weaker than gravity
as the electric field $F$ is stronger.
I consider it no further.

\section{The absence of Umklapp \label{S:UMKLAPP}}

A  lattice space-time  implies a cyclic energy-momentum space and Umklapp,
a violation of energy-momentum conservation by the amount of the momentum period.
$\3S$ quantum space-times like $\3Y$ are not lattices but have
non-commuting position variables and non-commuting momentum variables
with bounded discrete spectra,
neither periodic nor cyclic.
They add like non-commutative angular momenta $J_{m'm}$,
not commutative momenta.
For example,  a momentum componen $p$ obeys
\BEQ
-j\le J=p \le +j.
\EEQ
There is therefore no Umklapp.
The bounds have entropy 0, however,
requiring
temperature 0, and so
are still experimentally unattainable,
as when they were at infinity,
but now they can be approached as closely as experimental
resources allow, like absolute zero.

\section{Summary}
The indefinite Minkowski metric of space-time is
a singular limit of a neutral modularized Yang space-time metric,
which in turn 
derives from the Killing form of the basic cell.

Cartan  used Grassmann algebra as unifying language for differential geometry but could not fit the symmetric metric into it, 
and so he introduced an ad hoc tensor product as well.
His underlying manifold too is not defined today 
within a Grassmann algebra but within set theory.

In these $\3S$ models a multiordinal Grassmann algebra
with its Clifford operator algebra serves
as a universal language. 
The coordinate differentials are second-grade
Clifford elements.
Their metric form is now a fourth grade Clifford form,
automatically symmetric under the exchange of 
two second-grade forms from different cells.

Now Grassmann algebra supplemented 
with the operator of association is the common language not only for the tangent
structure but for the whole theory.
For one cell,
the Einstein metric tensor is the Killing metric tensor.

$\3S$ models based on Yang $\spin(5,1)$ or $\spin(3,3)$
predict a strong violation of the Heisenberg Indeterminacy Relations
between position and momentum,
allowing both their dispersions to be small at once,
when $|J|\ll \4N$ and
$i$ is disorganized.

I tentatively associate rank 1, 2, and 3 of $\3S$ with the first generation
neutrino space, lepton space, and fermion space of the Standard Model, each the Grassmann algebra of the preceding space;
and the quantized $i$, with an extra
spin operator $J_{65}$ like Yang.
This incorporates the three quark colors and 16 fermion flavors.
The relation between chirality and isospin is determined by the 
structure of the vacuum.
Space-time emerges from a coherent organization of the
 ranks from the spin rank 3 to the particle rank 5.
 

When the orbital Lie algebra
$\4a_{\0{orb}}$
is reformed to $\spin(3,3)$ (or $\spin(5,1)$),
all the orbital variables including $i$
become components of the quantified angular momentum tensor $J=J_{y'y}$ of rank 3 
($y,y'=1,\dots,6.)$ represented on rank 5.

%

The rank-4 fermion space $\3S(4)$ contains 
the rank-3 lepton space $4\1R$
and also three replicas of it, the quark $12\1R$.
That is why there are three colors.

Like set theory, 
the quantum architecture illustrated here 
is modular in the sense of Simon \cite{SIMON1962}.
 The spins 1/2 in odd quantum theories
naturally  have various dimensionalities.
This departs from the inspiring proposals of 
Penrose \cite{PENROSE1971} and Feynman \cite{FEYNMAN1941},
and others, but
brings their basic idea
closer to the working quantum theories of today.
Finite values for all observables is one immediate consequence:
Clifford elements have finite spectra.

Because the Heisenberg indeterminacy principle is so weakened,
it can no longer be excluded that gaugeons are pairs of 
odd quanta, though these odd quanta are not necessarily 
the ones able to
exist as free quanta.
 
I have described a kinematics, but to compute masses and cross sections a more detailed
dynamics must be set up and 
 solved.
 Transition amplitudes of an odd quantum theory are
 traces of Clifford polynomials of huge degree,
 easy to write and hard to compute.
 Brute-force computation at the microscopic level for a macroscopic process is
 beyond the scope of any conceivable artificial computer. 
  On the other hand, processes confined to
  some dozens of cells should be computable and might tell us something about very high energies;
  and some large but finite series can be summed in
  closed form.
  
  Feynman was concerned by the Umklapp phenomenon that arises
  in a lattice space-time,
  with large violations of particle energy-momentum conservation.
  Umklapp happens because a spacial lattice breaks Poincar\'e (or Yang) invariance badly at momenta high enough to resolve the lattice. 
 A quantum space-time with the Yang orbital algebra bends translational invariance slightly
  and
  has no Umklapp.
  
  The quantization of $i$ replaces the Heisenberg indeterminacy relation by
the spin indeterminacy relation.
In principle the difference is observable and fixes
the quantum of time.

\section{Acknowledgments}

I thank S. R. Finkelstein,  H. Saller, and Frank ``Tony" Smith, 
for discussions and information.
Dialogues with Tenzin Gyatso helped me
to shed the hypothesis of a permanent Natural Law \cite{FINKELSTEIN2004}.
The crucial importance of modularization dawned on me 
only after several of the annual Lindisfarne lectures of Lynn Margulis.
Presentation of this work was aided by FQXi and the Templeton Foundation.


\begin{thebibliography}{MM}

\bibitem{FINKELSTEIN2014}
D. R. Finkelstein.
Quantum field theory in quantum set algebra.
Submitted for publication in March, 2014.

\bibitem {SNYDER1947}
H. P. Snyder.
Quantized spacetime.
{\em Physical Review} 71:38 (1947)


\bibitem{YANG1947} 
C.N. Yang. On quantized space-time.
{\em Phys. Rev.} 72:874
(1947).

\bibitem{FEYNMAN1941}
R. P. Feynman.
Personal communication ca. 1961.
Feynman did this work ca. 1941.

\bibitem{PENROSE1971}
R. Penrose.
Angular momentum: an approach to combinatorial
space-time.
In T. Bastin (ed.),
{\em Quantum Theory and Beyond}\/,
151--180.
Cambridge 1971.
Penrose kindly shared much of this seminal work
with me ca 1960, long before publication.


\bibitem{SIMON1962}
H. Simon. The architecture of complexity.
{\it Proc. Amer. Philosophical Soc.} 106:6 (1962).

\bibitem{KOZO2010}
Symbiogenesis: A New Principle of Evolution.
B. M. Kozo-Polyansky (Author), L.  Margulis (Editor), V. Fet (Translator), Peter H. Raven (Introduction).
Harvard University Press (2010).

\bibitem{MARGULIS2002} L. Margulis and D. Sagan. 
{\em Acquiring Genomes: A Theory Of The Origin Of Species.} 
 (2002).


\bibitem{DIRAC1974}
P. A.M. Dirac.
{\em Spinors in Hilbert Space}\/.
Plenum, New York (1974)

\bibitem{WHEELER1983}
J. A. Wheeler. 
On recognizing `law without law'.
{\em American Journal of Physics} 51, 398-404 (1983).



\bibitem{WIGNER1967}
E. P. Wigner.	
{\em Symmetries and Reflections: Scientific Essays of Eugene P. Wigner}\/.
Indiana University Press, Bloomington
	(1967)
	

\bibitem{PALEV1977}
T. D. Palev.
Lie algebraical aspects of the quantum statistics. {U}nitary
quantization ({A}-quantization).
Joint Institute for Nuclear Research Preprint JINRE17-10550.
Dubna (1977).
hep-th/9705032.

\bibitem{PALEV2002}
T. D. Palev
and
J. Van der Jeugt.
Jacobson generators, Fock representations and statistics
of $sl(n+1)$.
{\em Journal of Mathematical Physics} 433, 850-3873 (2002).

\bibitem{FINKELSTEIN2013}
D. R. Finkelstein.
Palev statistics and the chronon.
In V. Debrev (ed.),
{\em Lie Theory and its Applications in Physics: IX International Workshop}\/.
Springer (2013).




\bibitem{FINKELSTEIN1955}
D. Finkelstein.  On relations between commutators.
{\em  Communications
in Pure and Applied Mathematics} 8:245-250 (1955)


	





\bibitem{LIEB1973}
E. H. Lieb. The classical limit of quantum spin systems. {\em Comm. Math. Phys.} 31:327-340
(1973).



 
\bibitem{SALLER1974}
H. Saller.
Gauge fields as bound states of subcanonical fermion fields.
{\it Il Nuovo Cimento A} Series 11, 24, 391 (1974).



\bibitem{TAVEL1965}
M.~Tavel, D.~Finkelstein, and S.~Schiminovich.
Weak and electromagnetic interactions in quaternion quantum
mechanics.
{\em Bull. Amer. Phys. Soc.} 9:435 (1965).



\bibitem{UNGER2007}
R. M. Unger.
{\em The Self Awakened: Pragmatism Unbound}
Harvard U. Press, Cambridge (2007).

\bibitem{FINKELSTEIN2004}
D. Finkelstein and A. Zajonc. Space, Time and the Quantum. Chapter 3 in
Arthur Zajonc (Editor),
{\em The New Physics and Cosmology: Dialogues with the Dalai Lama}\/. Mind and Life  (2004).
This describes the conference Mind and Life VII (1998).












%

 
%
 

%



\end{thebibliography}
\end{document}